\documentstyle[aps,multicol,prl]{revtex}
\begin{document}
\draft
\vspace{3cm}

\title{Magnetic circular dichroism in X-ray fluorescence 
of Heusler alloys at threshold excitation}

\author{M.V. Yablonskikh, V.I. Grebennikov, Yu.M. Yarmoshenko and E.Z. Kurmaev}
\address{Institute of Metal Physics, Russian Academy of Sciences-Ural Division,
620219 Yekaterinburg GSP-170, Russia} 
\author{S.M. Butorin, L.-C. Duda, C. Sothe, T. K$\ddot{a}$$\ddot{a}$mbre, M. Magnuson and J. Nordgren}
\address{Physics Department, Uppsala University,Box 530, S-75121 Uppsala, Sweden}
\author{S. Plogmann and M. Neumann}
\address{Universitat Osnabr$\ddot{u}$ck, Fachbereich Physik, D-49069 Osnabruck, Germany}

\maketitle

\begin{abstract}

The results of fluorescence measurements of magnetic circular dichroism (MCD) in Mn 
L$_2$,$_3$ X-ray emission and absorption for Heusler alloys NiMnSb and Co$_2$MnSb are presented. 
Very intense resonance Mn L$_3$ emission is found at the Mn 2p$_{3/2}$ threshold and is 
attributed to a peculiarity of the threshold excitation in materials with the half-metallic character of 
the electronic structure. A theoretical model for the description of resonance scattering of 
polarized x-rays is suggested. 
\end{abstract}

\begin{multicols}{2}

Heusler alloys with chemical composition  X$_2$MnZ or XMnZ (X=Cu,Ni,Co,Fe,Pd,Pt; 
Z=Al,Ga,In,Si,Sn,Sb) and L2$_1$ or C1$_b$ crystal structure are of great interest since their discovery 
in 1903 [1] because they are ternary intermetallic compounds with magnetic properties
that can be altered by changing the degree or type of chemical order [2]. First-principles spin-polarized 
calculations showed that the electronic structure of Heusler alloys has a metallic character for 
majority spin-electrons but an insulating character for minority spin-electrons which can induce 
the half-metallic ferromagnet state and band gap near the Fermi level[3].

It is supposed that X-ray fluorescence spectroscopy, performed with circularly polarized  
photons, could measure the spin polarization of the occupied density of states (DOS). This kind 
of dichroism in X-ray fluorescence closely related magnetic circular dichroism (MCD) in 
absorption [4] and is complementary to it. It is shown that a spin-polarized core hole may be used 
as a local site-specific spin detector for the valence states via appropriate valence-to-core dipole 
transitions. According to these expectations,the MCD effect can be observed without determining 
the polarization of emitted photons if the core holes are excited alternatively by the right- and 
left-handed circularly polarized radiation by using different sample magnetization directions. In 
this case the difference in the emitted radiation will closely reflect  the spin-resolved local 
DOS.

MCD in X-ray emission is predicted theoretically in Ref.5, based on fully
relativistic  spin-polarized band structure calculations for iron. Experimentally,
this effect is confirmed for  pure Fe in Ref [6] and then is used for the study of
the electronic structure of other magnetic systems as  Co and Ni [7], Rh , Rh$_{25}$Fe$_{75}$ and
Ni$_{45}$Fe$_{55}$ [8] and Fe-Co alloys [9]. In this letter we present the  results of study of
magnetic dichroism in Mn 2p X-ray emission and absorption for Heusler  alloys
NiMnSb and Co$_2$MnSb using energy-selective monochromatic excitation with circularly 
polarized X-rays.

The measurements were performed on beamline ID12B at the European Synchrotron 
Radiation Facility (ESRF). 
This beamline consists of a Dragon-like spherical
grating monochromator producing
some 83\% circularly polarized x-rays.
The extraordinarily low emittance of
the 6 GeV stored electron beam allowed us to refocus the x-ray beam that passes the
monochromator exit slit into a spot with dimensions of about 40 micron x 1 mm without 
excessive loss of intensity. The permanent magnet (NdFeB) devices were used to 
magnetize the sample for a fixed magnetization direction with H=0.2 T. The X-ray emission 
spectrometer consisted of an entrance 20 mkm slit, three spherical diffraction gratings, and a 
two-dimensional position-sensitive multichannel detector [10]. It was oriented with its optical 
axis perpendicular to the incident X-rays, in a vertical dispersive geometry and was operated 
with a 1200 lines/mm grating in second order of diffraction at a spectral resolution of 1.1 eV.
A single crystal of NiMnSb and a polycrystalline sample of Co$_2$MnSb were taken for X-ray 
fluorescence measurements. The samples were scraped in vacuum of 10$^{-6}$ torr before the 
measurements. 
Fig. 1 shows the MCD effect in X-ray absorption and emission of NiMnSb excited by the  right-
and left-polarized radiation. The Mn 2p X-ray absorption spectrum (XAS) measured in a 
total electron yield (TEY) mode shows dichroism (Fig. 1a)  at both  L$_3$ 
(E$_{exc}$=640.5 eV) and L$_2$ (E$_{exc}$=652.0 eV) thresholds. The MCD effect is stronger at the L$_3$ 
threshold than  that at L$_2$ threshold and a reason of this difference will be 
discussed later. The Mn 3d DOS is also presented.

Mn 2p X-ray emission spectra (XES) measured at the L$_3$ threshold (E$_{exc}$=640.5 eV), L$_2$ 
threshold (E$_{exc}$=652 eV) and far above threshold (E$_{exc}$=680 eV) show quite different fine 
structures. The Mn L$_3$ XES measured at E$_{exc}$=640.5 eV (Fig. 1b), which corresponds to
the  3d 4s$\to$2p$_{3/2}$ transition, has two subbands  A and B located at 637.0 and 640.5 eV, respectively. Both 
subbands show dichroism with the same sign as is found for the Mn L$_3$ XAS. The MCD effect reaches 
its maximum at emission energy E=640.5 eV which exactly corresponds to the Mn L$_3$ threshold. 
The double peak structure revealed in Mn 2p XES can be a result of superposition of spectra of 
normal emission (A) and elastic x-ray scattering, known as re-emission (B). The intensity of the B 
subband is found to be about 1.5 times higher than that of the A subband. The Fermi level
(estimated from Mn2p$_{3/2}$ core level photoemission [11]), is at the intensity minimum 
between these two subbands which means that the B subband corresponds to re-emission 
from unoccupied 3d states which are populated during near-threshold excitation of the Mn 2p$_{3/2}$ 
electron to the conduction band. X-ray re-emission is usually observed in spectra of 
insulators (see, for example, reference [12]) but in the case of a Heusler alloy, it  have been never seen with so high intensity.

In the Mn L$_3$ XES measured at the L$_2$ threshold excitation (E$_{exc}$=652 eV), we found great changes 
in the intensity distribution. The intensity of the peak A with an emission energy E=637.5 eV is decreased with 
respect to the spectrum measured at E$_{exc}$=640.5 eV and this peak is merged with the peak B forming a 
rather wide emission band centered at the emission energy E=638.3 eV. The main intensity is due to 
the Mn L$_2$ XES (3d4s$\to$2p$_{1/2}$ transition) which again shows two peaks (A' and B') located at emission 
energies of 648.2 and 652.0 eV (with Fermi level in between these peaks) corresponding 
to normal emission and re-emission, respectively. The intensity ratio of relative intensity I(B)/I(A) in Mn 
L$_2$ XES is found to be opposite with respect to that of the Mn L$_3$ XES. Dichroism of the Mn L$_3$ XES 
(A) at the L$_2$ threshold is found to be higher than that of the Mn L$_3$ XES (A) at the L$_3$ threshold 
whereas dichroism of the Mn L$_3$ (B) at the L$_2$ threshold is absent. 
The Mn L$_{2,3}$ XES measured far above the thresholds at E$_{exc}$=680 eV (Fig. 1d) shows 
dichroism with different sign for  L$_3$ and L$_2$ and relatively high ratio of 
I(L$_2$)/I((L$_3$)=0.92 which is higher than that for pure Mn (0.27). We have found also
anomalously high ratio I(L$_2$)/I((L$_3$)= 0.5-0.6 for that of Mn L$_{2,3}$ XES of Heusler alloys, measured with
  electron excitation [13].
Fig. 2 shows MCD in  Mn 2p X-ray emission for polycrystalline Co$_2$MnSb excited by 
right- and left-polarized radiation at E$_{exc}$=640.5, 644, 652 and 680 eV. As seen, 
the MCD behavior for Co$_2$MnSb is found to be almost the same as that for NiMnSb.
Therefore, one can conclude that experimentally observed MCD effects in near-threshold excited Mn 2p XES are 
the same for Heusler alloys with half-metallic character of electronic structure and can be 
discussed together. 

To interpret obtained results we have developed a theory for the description of X-ray transitions under 
threshold excitation. Here we will discuss only the main aspects of this theory. The details are  
given in [14]. The model is based on the calculated partial Mn 3d-density of states for NiMnSb [15] 
which has the following main features: (i) a strong exchange splitting (about 2 eV) which affects
the valence states distribution, (ii) spin-down states have a strong peak above the Fermi level (spin-up states 
are absent in this region and located below the Fermi level). We will show that this peak can be 
used as a trap for an excited core electron with a spin selectivity and responsible for the high intense 
re-emission peak in XES. Let us discuss the MCD effects in XES of Heusler alloys. Core states 
are split by the spin-orbital interaction for 2p$_{3/2}$ and 2p$_{1/2}$ with mostly parallel and anti-parallel 
direction of the orbital moment with respect to the spin moment of an electron. The spin direction is 
fixed by the spin-polarization of unoccupied states in which core electron are excited under 
absorption of x-ray quanta. Therefore, in such a way the part of states with orbital moment 
projections parallel (or anti-parallel) to the spin moment of atom are selected. The photons with 
right (photon spin +1) and left (photon spin -1) polarizations interact differently with electrons having a fixed 
projection of the orbital moment.This leads to dichroism. X-ray emission is a 
reverse process with respect to x-ray absorption and therefore also reveals dichroism. Using theory of 
the second order on electron-phonon interaction we have given the quantitative description of 
emission, re-emission and absorption within the  one-particle approach
In the case of resonance 2p$_{1/2}$ and 2p$_{3/2}$ excitations with right-hand photons (q=+1),
 when the photon spin is parallel to the spin moment 
of the solid, the emission intensities  are given by equations [14].
   This change of the dichroism sign can for differerent threshold excitations can
be explained within framework of the two-step process for the X-ray emission. At first, the transition of core 
level electron to the conduction band takes place which is characterized by dichroism of 
absorption and then a normal emission occurs (with an X-ray transition from the valence band to 
the vacancy on the core level) which is characterized by dichroism of emission. In the case of the 
near-threshold excitation, these both constituents of dichroism have different signs because spin-
polarizations of unoccupied and occupied states are different. In given case, dichroism of the first 
process (absorption) prevails. Under excitation far away from the threshold, the unoccupied 
states are unpolarized, $P^e$=0, and the result is determined by dichroism of the second step, 
i.e. emission.

	Going back to the re-emission peaks at L$_3$ and L$_2$ thresholds we need to note that the 
relative intensity of a first one is higher than a second one. Though the calculations predict the 
decrease in the ratio $I^{r}$/$I^{e}$ from 1.0 to 0.45 on going from L$_3$ to the L$_2$ threshold, 
the observed re-emission peak at L$_2$ threshold is even lower than predicted. It can be due to a shorter
lifetime of a L$_2$ core hole (compared to that of L$_3$ core hole) as a result of  L$_3$ $\to$ L$_2$ Auger 
transition. One can also expect that the field of a L$_2$ core hole with higher binding energy is 
screened stronger than that of a L$_3$ core hole. A value of the field directly determines the lifetime
 of the excited electron on the central atom and a probability of re-emission.

		In conclusion we have found MCD effects in X-ray emission and absorption of 
Heusler alloys which evidence about strong exchange splitting of Mn 3d-states with 
different spin projections. The observed giant re-emission peak in the Mn XES at the L$_3$ threshold  
is due to the half-metallic character of the electronic structure of Heusler alloys. The long lifetime
of the excited states (which is necessary for intense re-emission) is provided by 
the limitation of the radiation-less relaxation of excited electrons on the Fermi level and their delay on the atom by 
the field of the core hole.
 In Heusler alloys with large local magnetic moment there is a 
possibility of revealing of  such a mechanism of suppression of the relaxation which is based 
on strong exchange splitting, typical for half-metallic systems.
Both the existence of the energy gap in spin-down projection of Mn 3d-states
and weak hybridization of Mn 3d electrons with nearest neighborhood  conduce to the suppression of the 
excited electron relaxation. 
A theoretical model for the description of resonance scattering of polarized X-rays is suggested and the 
quantitative theory of X-ray emission MCD in terms of spin-polarized electron density of 
states and spin-orbital splitting of core levels is developed.

This work was supported by Russian Science Foundation for Fundamental Research 
(Project 96-15-96598 and 99-02-16266), NATO Linkage Grant (HTECH.LG 971222), Swedish Natural Science Research Council (NFR) and Goran Gustavsson Foundation in Natural Sciences and Medicine,Deutsche Forschungsgemeinschaft {DFG project Br 1184/4} and Bundesministerium fur Bildung und Forshung (BMFB 05SB8MPB8). Technical assistance of ESRF staff is gratefully acknowledget. The single crystal of NiMnSb was supplied by Ch. Hordequien (CNRS) and polycrystaline Co$_2$MnSb was supplied by Elena I. Shreder
 (Institute of Metal Physics, Russian Academy of Sciences-Ural Division)

\end{multicols}
Figures Caption:

Fig. 1  Excitation energy dependence of the Mn L$_2$,$_3$ XES and Mn 2p XAS of NiMnSb.
Mn 2p TEY spectrum (a) and  XES spectra (b), (c), (d) ,excited with parallel aligment of photon spin and spin moment 
(filled squares) and antiparallel aligment (open circles). 
Mn 2p$_{3/2}$ and Mn2p$_{1/2}$ binding energies relative to Fermi level are indicated as E$_f$(L$_3$) and E$_f$(L$_2$),respectively
Energy resolution is 1.1 eV
The Mn 3d DOS [15] is shown at insertion at Fig 1.b

Fig. 2 Excitation energy dependence of Mn L$_2$,$_3$ XES of Co$_2$MnSb.
Spectra (a), (b), (c), (d) excited with  parallel aligment of photon spin and spin moment (filled squares),
and  antiparallel aligment (open circles). Corresponding excitation energies are shown right side of frame.
Mn 2p$_{3/2}$ and Mn2p$_{1/2}$ binding energies relative to Fermi level indicated as E$_f$(L$_3$) and E$_f$(L$_2$),
respectively.Energy resolution is 1.1 eV.

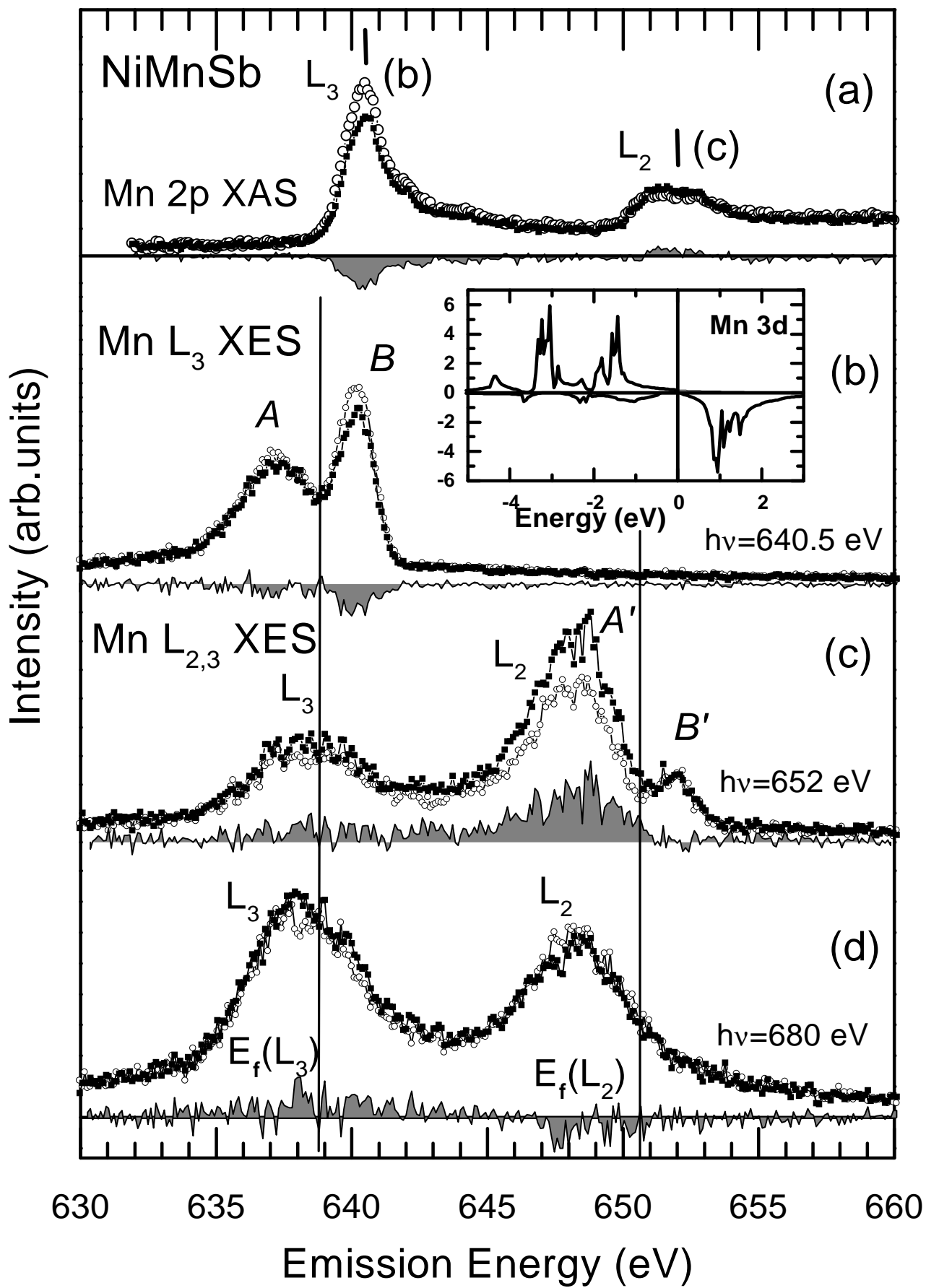
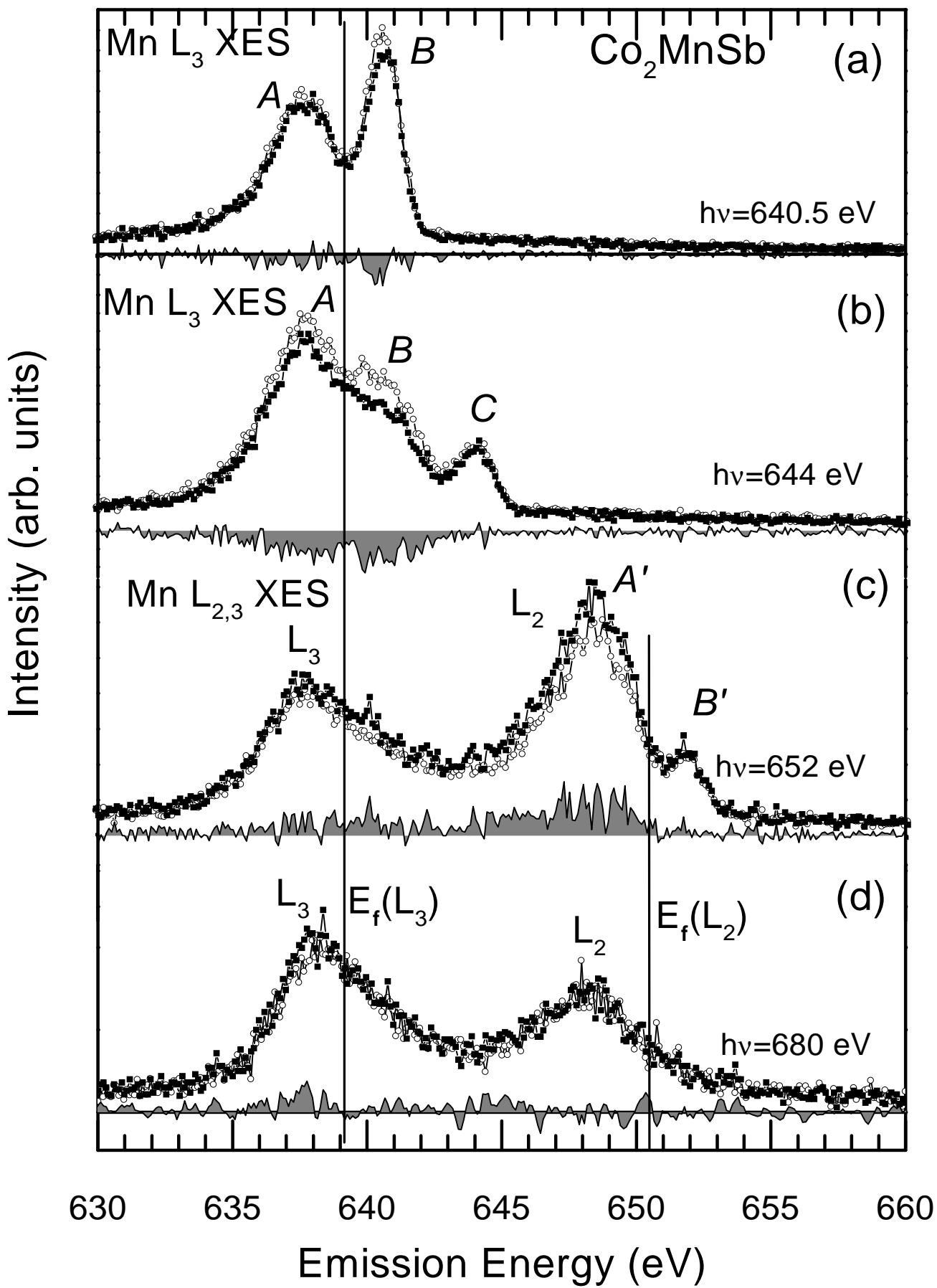
\end{document}